# Investigating Benefits and Limitations of Migrating to a Micro-Frontends Architecture

Fabio Antunes
Tecgraf / PUC-Rio
Rio de Janeiro, Brazil
fabioagpp@tecgraf.puc-rio.br

Maria Julia Dias de Lima
Tecgraf / PUC-Rio
Rio de Janeiro, Brazil
mjulia@tecgraf.puc-rio.br

Marco Antônio Pereira Araújo
UFJF / IF Sudeste MG
Juiz de Fora, Brazil
marcoaparaujo@gmail.com

Davide Taibi
University of Oulu
Oulu, Finland
davide.taibi@oulu.fi

Marcos Kalinowski
ExACTa / PUC-Rio
Rio de Janeiro, Brazil
kalinowski@inf.puc-rio.br

## ABSTRACT

**[Context]** The adoption of micro-frontends architectures has gained traction as a promising approach to enhance modularity, scalability, and maintainability of web applications. **[Goal]** The primary aim of this research is to investigate the benefits and limitations of migrating a real-world application to a micro-frontends architecture from the perspective of the developers. **[Method]** Based on the action research approach, after diagnosis and planning, we applied an intervention of migrating the target web application to a micro-frontends architecture. Thereafter, the migration was evaluated in a workshop involving the remaining developers responsible for maintaining the application. During the workshop, these developers were presented with the migrated architecture, conducted a simple maintenance task, discussed benefits and limitations in a focus group to gather insights, and answered a questionnaire on the acceptance of the technology. **[Results]** Developers' perceptions gathered during the focus group reinforce the benefits and limitations reported in the literature. Key benefits included enhanced flexibility in technology choices, scalability of development teams, and gradual migration of technologies. However, the increased complexity of the architecture raised concerns among developers, particularly in dependency and environment management, debugging, and integration testing. **[Conclusions]** While micro-frontends represent a promising technology, unresolved issues still limit their broader applicability. Developers generally perceived the architecture as useful and moderately easy to use but hesitated to adopt it.

## KEYWORDS

Micro-frontends, Benefits and Limitations

## 1 INTRODUCTION

As time passes and new features are added to applications, both the frontend and the backend code bases grow [18]. Within a monolithic structure, this expansion may lead to a slowdown in development due to conflicting code management, an increasing number of dependencies, increased coupling, and the sheer amount of code developers need to be familiarized with to maintain the application.

Micro-frontends have garnered attention for their potential to address these challenges by enhancing application modularity, scalability, and maintainability [9]. It is defined as "an architectural style where independently deliverable frontend applications are composed into a greater whole" [9]. The adoption of micro-frontends aims to achieve the same benefits that microservices bring on the backend side in the frontend development of applications.

Nonetheless, the micro-frontend architecture is still in its early years, and few studies have investigated it. Peltonen *et al.* [18] mapped benefits and issues in a multivocal literature review, analyzing sources from academic and industry literature. Their review could be enhanced by incorporating case studies and reports detailing practical experiences related to the architectural transition. Such insights would contribute to the consolidation of knowledge concerning the potential benefits and limitations.

This paper aims to explore the benefits and issues of the micro-frontends architecture migration in an industry project. Therefore, an experienced developer of the organization (first author), together with the project manager who has worked on the project for over ten years and has a deep understanding of its architecture (second author) and software engineering and micro-frontends researchers, applied an investigation strategy based on action research to transition a monolithic web application to a micro-frontend architecture and assess the effects from the remaining developers' point of view.

The objective of this research can be defined using the goal-question-metric (GQM) [4] goal definition template as follows.

*Analyze* micro-frontends architecture when compared to a monolithic architecture *for the purpose of* characterizing *with respect to* benefits, issues, and technology acceptance *from the point of view of* the developers *in the context of* scrum teams experienced in the development of the monolithic version of the web application chosen as the subject of this study. The following research questions (RQs) were formulated based on the benefits and issues of the micro-frontends architecture raised by Peltonen *et al.* [18].

**RQ1. What are the benefits and issues of the micro-frontends architecture when compared to a monolithic architecture?** This RQ explores the influence of the architecture on the ease of the development process. It aims to evaluate both the benefits and challenges identified by Peltonen *et al.* [18]. Additionally, we intend to evaluate the potential enhancement in the reuse of application components that it might offer [13].

**RQ2. What is the acceptance of the micro-frontends architecture when compared to the monolithic architecture?** This RQ investigates how the developers accept the technology by using the Technology Acceptance Model (TAM) [21] to assess the perceived usefulness, ease of use, and intention to adopt.

Our methodology phases are based on action research [19] to achieve the stated goals and effectively address the research questions. Action research, known for its iterative and participatory nature, emphasizes collaboration between researchers and practitioners to enact meaningful change within a specific context [19]. This methodology aligns with our research objectives by facilitating engagement with developers involved in the industry project [7].

The methodology comprises four phases: diagnosis, planning, intervention, and evaluation of intervention results. In this study, the initial diagnosis identified a challenge of two teams struggling to collaborate on the same web application. The intervention comprised migrating the application to a micro-frontends architecture. To evaluate the intervention, a workshop was conducted to introduce the architecture to the remaining developers and engage them in a practical exercise. Subsequently, their feedback on its benefits and limitations was assessed via a focus group [10]. Due to practical constraints, we did not employ a completely participatory approach involving all employees in all steps but involved them within our possibilities. The research team comprised an experienced developer of the organization (first author), the project manager (second author), and three software engineering researchers, including one micro-frontend expert. During the diagnosis, besides having access to all artifacts, we interviewed the remaining project employees. During the evaluation, we also had the remaining employees in a workshop where we presented the intervention and had them participate in the practical exercise and the focus group discussions.

The contributions of this paper comprise the architectural decision rationale of the intervention and the results of the discussions of benefits and limitations with the practitioners. While developers generally perceived the architecture as useful and moderately easy to use, they also exhibited hesitance toward adopting it, mainly due to difficulties related to dependency and environment management, debugging, and integration testing.

## 2 BACKGROUND AND RELATED WORK
### 2.1 Micro-frontends

Micro-frontends architecture divides web applications into smaller, independent frontends that collaborate to create the complete application [18]. Each micro-frontend operates with its own codebase, fostering loose coupling and enabling autonomous execution, development, and deployment. Four pivotal technical decisions at the project's outset dictate subsequent development: how the application will be split, how its components will be assembled, how the application routes between pages, and how the micro-frontends communicate with each other [18, 20].

*Splitting the Frontend*. There are two options available for splitting the frontend: vertical and horizontal. In a vertical split, only one micro-frontends is displayed at a time, often segmented based on business domains [13]. On the other hand, a horizontal split simultaneously displays multiple interface sections.

*Composition Techniques*. Three primary composition techniques exist: client-side, edge-side, and server-side [13]. Client-side composition loads micro-frontends in the browser, while edge-side assembly occurs at the Content Delivery Network (CDN) level, employing technologies like Edge Side Include (ESI). Server-side composition can occur during compile or runtime, with methods like server-side includes (SSI) or ESI [18].

*Routing*. There are three possible routing methods that vary based on composition type [13]. For server-side composition, it is possible to have the logic that determines the route defined on the server. On a client-side composition, routing can be defined by the application shell by evaluating its state. Alternatively, independent of the composition type, it is possible to manage routing on the CDN, employing transclusion technologies such as Lambda@Edge [13].

*Communication*. Data communication between micro-frontends can use built-in browser structures like Web Storage or query strings [11, 15]. In the case of a client-side composition, where interactions between micro-frontends are more complex, it might be necessary to have a publish-subscribe pattern with an event bus that can be observed by all micro-frontends [13].

### 2.2 Benefits and Issues of Micro-frontends

Based on previous research, micro-frontends architecture may offer several benefits (B) and issues (I) [18]. Given that our research is related to these benefits and issues, we summarize them hereafter.

- **B1 - Support for different technologies:** Enables coexistence of diverse technologies, facilitating flexible development choices.
- **B2 - Autonomous cross-functional teams:** Empowers teams to work independently on application parts, enhancing agility and ownership.
- **B3 - Independent development, deployment, and managing:** Decouples modules, enabling parallel work without coordination overhead.
- **B4 - Better testability:** Changes in one micro-frontend have limited impact, speeding up testing and deployment.
- **B5 - Improved fault isolation:** Failures in one module do not affect the entire application, enhancing resilience.
- **B6 - Highly scalable:** Easy addition of new teams and tasks, fostering parallel development.
- **B7 - Faster onboarding:** New team members require less time to understand and contribute due to compartmentalization.
- **B8 - Fast initial load:** On-demand loading of modules speeds up application setup.
- **B9 - Improved performance:** Performance degradation in one part minimally affects the overall frontend.
- **B10 - Future proof:** Easy adoption of new technologies and abandonment of outdated ones due to decoupled structure.
- **I1 - Increased payload size:** Multiple frameworks lead to increased data fetching, impacting performance.
- **I2 - Code duplication and I3 - Shared dependencies:** Replicated dependencies result in increased complexity and potential redundancy.
- **I4 - UX consistency:** Achieving consistency across micro-frontends is challenging due to team independence.
- **I5 - Monitoring complexity:** Tracking bugs across micro-frontends and communication layers is complex.
- **I6 - Increased complexity:** Architecture complexity requires substantial analysis and decision-making.

- **I7 - Governance challenges:** Collaboration and alignment across teams pose governance challenges.
- **I8 - Islands of knowledge:** Isolation between teams can lead to knowledge silos and duplicated efforts.
- **I9 - Environment differences:** Discrepancies between environments may lead to unexpected behavior in production.
- **I10 - Higher risk in updates:** Reduced interval between development and release increases the risk of runtime errors.
- **I11 - Accessibility challenges:** Embedding iframes may pose accessibility issues, requiring careful consideration.

## 2.3 Related Work

Peltonen *et al.* [18] conducted a multivocal literature review on micro-frontends architecture, which guided the theoretical foundation of this paper. Taibi *et al.* [20] provided insights into principles, implementations, and pitfalls of micro-frontends, informing technical decisions of this study.

Harms *et al.* [8] experimented with different frontend architectures connected to microservices backends. Their findings highlighted good testability and modifiability in micro-frontends, with variations in performance and UI consistency.

Mena *et al.* [12] developed a multi-platform web application using micro-frontends and microservices to maintain user experience across diverse devices. While their article's main focus was the application's requirements, they mention using micro-frontends to isolate the development of the different components and avoid conflicts within the codebase.

Pavlenko *et al.* [17] detailed the development of a React single-page application using micro-frontends. The article reported that the architecture presents an increased complexity and suitability for applications with significant front-end logic and larger teams.

Männistö *et al.* [14] migrated from a monolithic application to a micro-frontend architecture using a frameworkless approach with Web Components [5]. Conducted on a small organization with few developers, their study indicates that even small teams can benefit from micro-frontends, particularly in terms of configurability. They also highlight that the frameworkless approach can be a feasible alternative, as it avoids the issues typically induced by frameworks.

## 3 RESEARCH STRATEGY

The research strategy employed in this work followed the conventional structure of action research, which includes four phases: diagnostic, planning, intervention, and evaluation [19]. In line with ethical guidelines, we explained the purpose and procedures of our study, communicated the risks and benefits of participating, and gathered the consent of the participants, explaining to them the anonymity of the data and their confidentiality[3].

## 3.1 Context

This study focuses on a system developed by a research institute for a large oil and gas company. Since 2001, the system has served as a support platform for constructing and integrating geophysical processing applications and orchestrating in-house algorithms' workflows on high-performance computing infrastructures. Over the years, multiple teams have contributed to the system's development, underscoring the importance of a flexible and decoupled architecture for the evolution and integration of new applications. Gradually, the system has been transitioning to a web architecture to adapt to changing technological landscapes.

The application that is the focus of our intervention supports geophysicists, who execute algorithm workflows following a sequence of phases that automatically configure most of these algorithms' parameters. The phases are built into users' projects using a pre-configured hierarchical model within the applications. The pre-configured phases model organizes the possible sequence of paths that the user can follow and establishes the input and output dependencies between the algorithm workflows of each phase.

The application consists of two pages: An initial **project-page** enables users to maintain project information, as well as navigate to the project's corresponding algorithm page; An **algorithm-page** houses a graphical representation of the different sequences of algorithms ran for that project in a tree form. The root node symbolizes the project and each subsequent node represents a phase of the algorithmic sequence. Users can select parameters for executing the algorithm of a phase multiple times. Being satisfied with the outcomes, geophysicists can generate child nodes, progressing the sequence of algorithms. The algorithm-page also allows users to edit project parameters by interacting with the root node.

The participants include developers from two scrum teams familiar with the subject application, totaling 10 professionals. Their software development experience spans four to twenty years, with two to six years in frontend web application development.

## 3.2 Diagnostic

This stage's purpose is to explore the research field, stakeholders, and their expectations in order to identify primary causes and circumstances faced by the organization [7]. Diagnostics was split into an analysis of the system and an interview with the developers.

*3.2.1 System Analysis.* The application under analysis was developed by one of the teams while they were still gaining experience in web development and learning to use a specific frontend framework (Angular 12). The first author, who is employed at the institute and actively involved in the project, led the analysis conducted in this step, with support from the project manager and the academic advisor. Being part of the development team, he had access to all necessary artifacts, ensuring comprehensive data collection. The project manager's long-term involvement and the academic advisor's expertise in software architecture contributed to a thorough and robust analysis.

Upon analyzing the project structure and source code, it became clear that each page's components had a high degree of interaction with each other. Some event-driven structures had already been put in place to reduce the coupling between components. However, the pages still require a considerable amount of setup and coordination in order for everything to work according to specifications. In addition, all the components share some common model objects, which means an additional shared dependency and that a new application would require it to use the exact same models in order to reuse the components.

The analysis indicated that a decoupled architecture would be desirable, enabling development teams to work simultaneously and efficiently on the same application. The goal of this architecture would be to gain agility through the composition of components that are common to these applications.

*3.2.2 Interview.* The interview consisted of a one-on-one discussion between the researcher and developers from both teams. Due to the project's schedule, seven developers were available for this step. It followed a questionnaire [2] created based on the researcher's observations during the system analysis. Its intent was to first raise any concerns the developers could spontaneously recall without external influence and then get their opinions on scaling up the development teams and improving code reusability. Key points raised during the diagnostic interviews include:

- Issues with the organization of a common library, slow test pipeline, and replicated code in end-to-end test suites.
- Challenges in adding team members included insufficient documentation, time required for understanding the application, and coordination difficulties for parallel team work.
- Division of application into separate repositories was viewed positively for limiting the scope, facilitating learning, and reducing maintenance efforts. However, concerns were raised about increased complexity and potential conflicts.
- Reusability of current components in new applications was perceived as feasible but requiring improvements in generality, modularity, and communication between components.
- Centralization of data was noted for reducing divergent information but increasing component coupling. Distributing data management responsibilities between components was seen as potentially reducing coupling but requiring caution for consistency and additional testing.

## 3.3 Planning

Once the need to improve the capacity of increasing the number of developers working in parallel on the same application with minimal code conflict was established, a search in the literature about micro-frontends was conducted. According to Peltonen *et al.* [18], this architecture is naturally loosely coupled, so it allows for easier onboarding of new teams and independent development of the different parts of the interface.

After reviewing the available literature, taking into consideration the context of the project and the benefits and issues raised by Peltonen *et al.* [18], it was decided to apply micro-frontends architecture to the application and analyze how it would compare to the team's past experience. The architecture would possibly solve the team's current inability to work together. Its loose coupling would also help to fulfill the foreseen requirement of making application replicas based on reusing different parts. To plan the migration, we analyzed the reported benefits and issues in the context of the system to be migrated. This rationale follows.

*3.3.1 Micro-frontends Benefits Analysis.* Micro-frontends offer several advantages outlined by Peltonen *et al.* [18], including the ability for multiple teams to work independently (**B2**, **B3**). This aligns with the system's need to reduce code conflicts and complexities, allowing teams to focus on specific parts of the application. The loose coupling of micro-frontends (**B6**) facilitates reuse and reduces redundant work. Additionally, it addresses knowledge dispersion (**B7**) by enabling swift comprehension of unfamiliar system parts.

Improvements in testing (**B4**) are anticipated, with smaller libraries and fewer tests. Although diverse technologies offer allure (**B1**), sticking to a singular framework streamlines integration. Future transitions to diverse technologies may require additional work (**B10**). Some benefits such as fault isolation (**B5**), fast initial load (**B8**), and improved performance (**B9**) may not be immediately apparent because the web application is embedded on a desktop-based system, but could become significant as it transitions to the web.

*3.3.2 Micro-frontends Issues Analysis.* Several challenges identified by Peltonen *et al.* [18] can be addressed through strategic planning in micro-frontends. Issues like "Increased payload size" (**I1**), "Code Duplication" (**I2**), and "Shared Dependencies" (**I3**) stem from diverse technologies across micro-frontends. Adopting the same framework through all libraries and micro-frontends would reduce the amount of dependencies. Additionally employing appropriate composition technology can further mitigate these issues by avoiding the same library to be loaded multiple times.

Addressing the issue of "UX consistency" (**I4**), a common library housing UI components and stylesheets emerge as a potential solution. However, this approach presents its own set of challenges. First, the possibility of conflicts arises as multiple teams may concurrently contribute to this shared resource. Second, any modifications to this library can inadvertently impact multiple micro-frontends. To strike a balance, teams must commit to following semantic versioning practices, thereby managing and mitigating these concerns.

Existing measures, such as an automated build and test pipeline linked to a repository manager, help to alleviate the "Increased level of complexity" (**I6**). Additionally, the presence of multiple homologation environments designed to simulate the production environment reduces the likelihood of bugs stemming from "Environment differences" (**I9**) and mitigates the "Higher risk when releasing updates" (**I10**) issue. Furthermore, potential "Governance" (**I7**) challenges are mitigated by the presence of a shared Project Owner overseeing both teams. While the new architecture may amplify these challenges, its overall impact is expected to remain minimal due to these preexisting safeguards.

The challenge of "Islands of Knowledge" (**I8**) is already prevalent in the current operational setup, given that teams work on distinct applications. In this regard, the adoption of smaller, more manageable projects aligned with the micro-frontends approach can potentially facilitate "Faster Onboarding" (**B7**) and help mitigate this problem. "Accessibility challenges" (**I11**) are less pertinent due to internal use, eliminating stringent accessibility requirements associated with public-facing applications.

## 3.4 Intervention

The existing monolithic web application was restructured into smaller, self-contained components, facilitating micro-frontends integration within the application shell. This horizontal split decision was driven by the shared component responsible for project updates and the promotion of common concepts for future reuse. Due to the team's expertise, all micro-frontends continued to use the Angular framework.

While the applications could potentially bundle all micro-frontends together at compile time, the decision was made to experiment with separate access via different URLs. The composition is performed on the client side using Module Federation, a technology already integrated into the application's bundler (webpack [16]). This choice facilitates the definition of shared libraries, loading them only once, and benefits from an existing Angular library that leverages Module Federation for loading separately compiled and deployed code [1].

The selected composition technology influences page routing within the application shell. Angular Architect's library enables association of remote modules to router URLs, enabling logic for micro-frontend display based on emitted events.

Communication between micro-frontends is facilitated by injecting an event emitter into each one of them via an Angular service. Additionally, direct access to the application's second page is enabled through URL-encoded variables.

To mitigate risks associated with the changes, all proposed modifications were implemented within separate repositories, distinct from those used in production. This approach allowed us to assess the intervention's impact without committing to changes. If approved, the adjustments can be integrated into the project; otherwise, they can be discarded without necessitating a rollback.

The extracted micro-frontends and created auxiliary libraries are briefly described hereafter.

*3.4.1 Micro-frontends.* The migrated application was divided into the following 5 micro-frontends, delimited by red rectangles in Figure 1 and Figure 2.

The **project-selection-microfrontend** (Figure 1 left side) displays a list of project cards and a button to create new projects. This micro-frontend is always visible on the first page. The **project-info-microfrontend** (Figure 1 right side) features three tabs for editing project data.

The **pipeline-microfrontend** (Figure 2 upper left) shows a tree-like graph of phase sequences created by the user and is always visible on the second page. The **phase-parametrization-microfrontend** (Figure 2 right side) allows users to parameterize and execute selected phases. Its fields adjust based on phase type and parent parameters. Finally, the **execution-table-microfrontend** (Figure 2 lower left) lists all executions for the selected phase. Both these micro-frontends are visible only on the second page when a phase is selected.

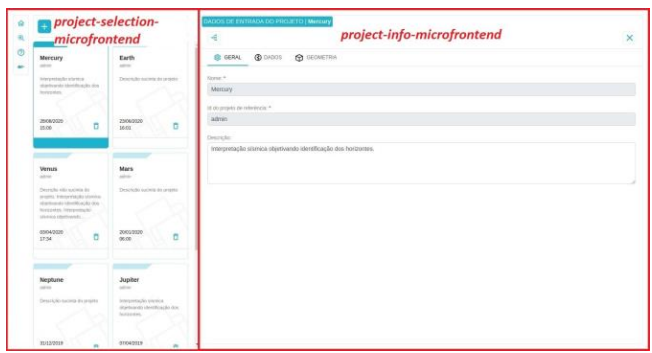

**Figure 1: Application: First page**

Based on our proposal for micro-frontends, we determined that implementing the **execution-table-microfrontend** was unnecessary because it primarily serves as a view-only component. Its functionalities were adequately covered by other components, offering no additional benefits.

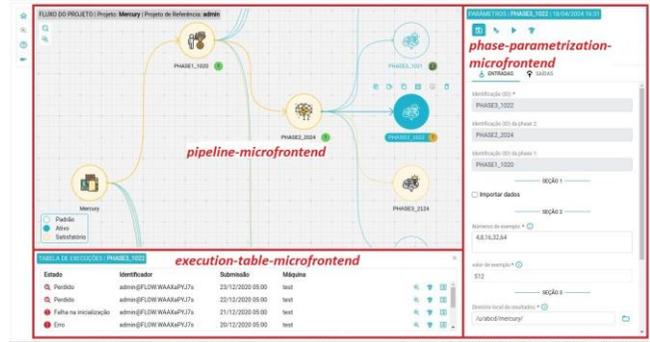

**Figure 2: Application: Second page**

*3.4.2 Auxiliary Libraries.* In addition to the micro-frontends, several libraries were created to establish a shared code base for integrating the application. Figure 3 illustrates the architecture, delineating the dependencies and development dependencies between the shell application, micro-frontends, and libraries. The diagram reveals two distinct categories for the auxiliary libraries (common library and feature library).

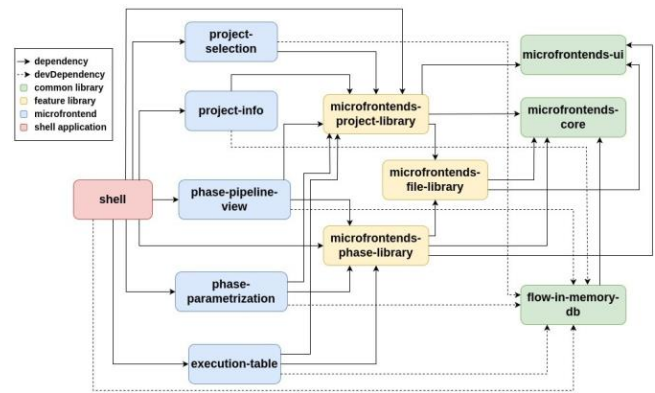

**Figure 3: Library breakdown**

Common libraries, imported by all micro-frontends, aggregate services and components aiding integration. **microfrontends-core** serves as the main library, defining services for environment properties and session variables. **microfrontends-ui** maintains UX consistency across micro-frontends. **flow-in-memory-db** simulates the backend, exclusively for development.

Feature libraries categorize application data into projects, files, and phases. Corresponding libraries consist of models, utility components, and services. The **feature service** provides access to and modification of data. Services' methods initiate authorization requests, interact with the backend, and update data. Careful handling of feature libraries is crucial to avoid breaking retro-compatibility.

## 3.5 Evaluation Strategy

The evaluation seeks to discuss the benefits, issues, and technology acceptance of the migrated architecture (intervention) with the developers. To achieve this, the developers engaged in a workshop in which the migrated architecture was presented, briefly used in practice, and discussed. The workshop was structured into the following three phases:

*3.5.1 Presentation Phase (45 minutes).* During this phase, the researcher provided participants with an overview of micro-frontends, explaining fundamental concepts, benefits, and challenges associated with the architecture. This sets the stage for participants to grasp the theoretical underpinnings of micro-frontends.

Then, the changes to migrate the application's software architecture to micro-frontends were presented. Technical decisions regarding the four micro-frontends principles were outlined. It was also highlighted how the adoption of micro-frontends was intended to address specific pain points and enhance the software's modularity and scalability.

*3.5.2 Practical Application Phase (45 minutes).* Participants were assigned the task of adding a new form field to the **project-info-microfrontend**. This exercise simulated a real-life maintenance task. The time for completion was 30 minutes, followed by a presentation of a potential solution to address any remaining doubts or incomplete tasks. In order to enable participants to complete the task in the allotted time, they were provided with an updated version of the library containing the models that should be updated and the shell application that contained the mocked backend.

*3.5.3 Focus Group Discussion Phase (60 minutes).* Following the practical exercise, a focus group discussion ensued, allowing each participant to voice opinions on the benefits and challenges of the architecture delineated in the literature review by Peltonen *et al.* [18]. This forum provided a platform for participants to share their perspectives, experiences, and concerns regarding the migrated micro-frontends architecture.

Additionally, each participant completed a characterization form to furnish background information, alongside a Technology Acceptance Model questionnaire[21] to assess their willingness to adopt the new architecture. The next section provides a more detailed view of this phase and its results.

## 4 EVALUATION

A focus group was selected as an evaluation method for its suitability to gather practitioner insights [10]. It fosters open discussions, allowing participants to share diverse viewpoints and uncover nuanced perspectives. As a result, we collected comprehensive feedback on the architecture and generated new discussion points. A repository was created containing all the artifacts of both the workshop and the focus group [2].

## 4.1 Focus Group Planning

The focus group centered on the benefits and issues uncovered by Peltonen *et al.* [18] and was conducted virtually via Zoom, using a Miro board for discussion. The board consisted of 21 statements related to the issues and benefits, each rated by developers on a Likert scale ranging from 1 (strongly disagree) to 5 (strongly agree). Organized into lanes, each statement had six columns (see Figure 4). Participants, distinguished by assigned colors, affixed post-its beneath their respective grades and provided justifications. This was followed by a designated period for verbal discussion. The post-it justifications and complete transcriptions of the discussions are available in our open science repository [2]).

The only issue that wasn't treated was **I11** since the software does not have accessibility requirements. Instead, we added a question (**Q11**) that focused on the potential reusability benefit mentioned by Mezzalira in his book [13].

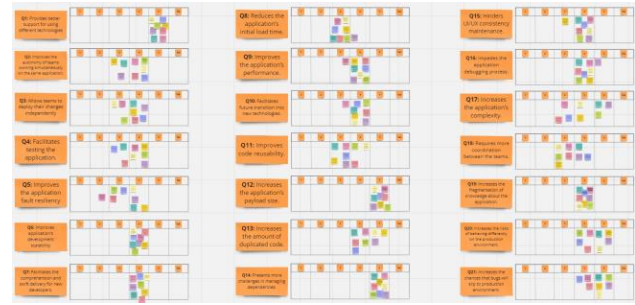

**Figure 4: Focus Group board after discussion (online at [2])**

Initially scheduled to last 60 minutes with a possible 30-minute extension, the focus group surpassed the anticipated duration. Given that participants had only covered half the board by the scheduled conclusion, a second session was arranged for the next day, with post-it assignments made beforehand to ensure timely completion.

Data from the focus group was captured through two methods. Firstly, participants provided justifications on the board. Secondly, the discussions were recorded, transcribed, and translated into English. Additionally, participants filled out a Technology Acceptance Model (TAM) questionnaire after the sessions.

## 4.2 Focus Group Results: Benefits and Issues

In this subsection, we answer RQ1 by presenting the results for each benefit and issue discussed in the focus group and offer a synthesis of the developers' opinions. Figure 4 illustrates the final state of the board after the discussion. Figure 5 presents a consolidated view of the final ratings for ease of interpretation.

**Q1 - Provides better support for using different technologies. (B1)** Participants unanimously agree that micro-frontends facilitate the use of diverse technologies within a single application, allowing for tailored solutions to address specific challenges effectively. However, they acknowledge the need for careful management to mitigate maintenance costs associated with utilizing multiple technologies.

**Q2 - Improves the autonomy of teams working simultaneously on the same application. (B2)** While opinions are divided, there is consensus that micro-frontends can enhance team autonomy to some extent, although effective alignment and communication remain crucial for smooth collaboration.

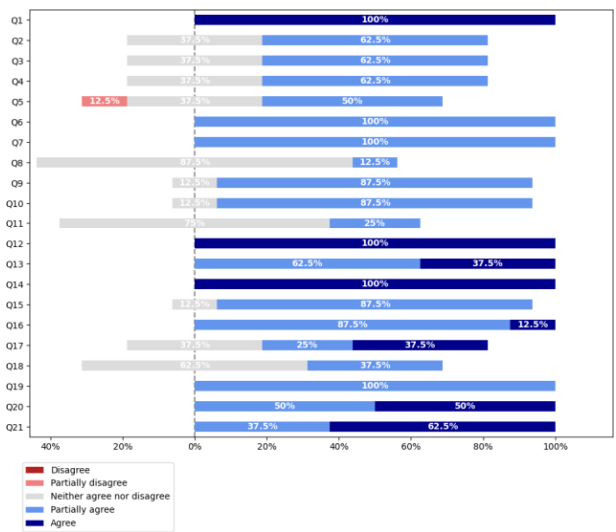

Figure 5: Focus Group results

**Q3 - Allows teams to deploy their changes independently. (B3)** Similarly, views on the ability of micro-frontends to enable independent deployment of changes vary, with concerns raised about the need for alignment between teams to avoid inconsistencies.

**Q4 - Facilitates testing the application. (B4)** There's a mixed perspective on whether micro-frontends facilitate testing the application. While some believe that it simplifies unit testing and improves test scope definition, others express concerns about potential complexities, such as code repetition in component tests and difficulties in integration testing and maintenance.

**Q5 - Improves the application's fault resiliency. (B5)** The discussion on whether micro-frontends improve fault resiliency reveals varying perspectives, with considerations about the impact of failures and the dependence on application division granularity.

**Q6 - Improves application's development scalability. (B6)** While participants agree that micro-frontends have the potential to enhance application development scalability, they highlight challenges related to component communication and team division.

**Q7 - Facilitates the comprehension and swift delivery for new developers. (B7)** All the participants partially agree that micro-frontends can aid in new developers' comprehension and help their integration into projects. The focused development scope within each micro-frontend enables swift onboarding. However, setting up the development environment and comprehending the integration of all components, including the shell and micro-frontends, may pose challenges.

**Q8 - Reduces the application's initial load time. (B8)** While there is recognition that the chosen approach for application composition, the division of the application, and the architecture play crucial roles, developers don't think the architecture necessarily improve initial load time. Responses highlight factors such as dependency duplication, the potential for improved loading through module loading, and the possibility of making parts of the application available before all components are fully loaded.

**Q9 - Improves the application's performance. (B9)** In general, they think that micro-frontends can boost performance by enabling different parts of the application to load independently, enhancing user experience and providing better control over loading. However, developers stress that their effectiveness in improving performance depends on factors like architectural design, division strategies, and scalability measures.

**Q10 - Facilitates future transition into new technologies. (B10)** They believe that micro-frontends offer advantages in terms of flexibility and gradual migration, simplifying the transition to new technologies. However, developers also highlight challenges such as the need to replicate changes across multiple submodules, complexity in integrating different parts, and the risk of compatibility issues with common libraries.

**Q11 - Improves code reusability.** Three developers partially agree that micro-frontends might offer potential for code reuse, especially if the same micro-frontend can be utilized across different applications, while four neither agree nor disagree. However, concerns have been raised about the effectiveness of code reuse within the application itself, with some developers expressing uncertainty or skepticism about whether micro-frontends truly enhance code reusability compared to monolithic architectures. Developers highlight that the architecture's gain in reusability depends on the choice of split and may still require customization to adapt the micro-frontend to its new use.

**Q12 - Increases the application's payload size. (I1)** There is a consensus among the developers that micro-frontends can lead to an increase in payload size. This might be due to various factors, such as duplicated dependencies, code, and components across different micro-frontends. Developers express concerns about managing and mitigating the duplication of dependencies and the potential impact on overall application performance. While some acknowledge the possibility of mitigating this issue through strategies like lazy-loading and caching, there is an overall recognition that micro-frontends introduce challenges in minimizing payload size.

**Q13 - Increases the amount of duplicated code. (I2)** Developers express concerns about code duplication, particularly with scripts, configuration files, deployment-related code, and backend data processing logic. They attribute this duplication to independent component development by different teams and the limited awareness or visibility of existing code across various parts of the application. Some suggest it's possible to place it in a common library. However, they also acknowledge the additional effort required.

**Q14 - Presents more challenges in managing dependencies. (I3)** There are notable concerns about the complexity and added management burden linked to dependencies in micro-frontends versus monolithic architectures. Developers point out challenges like conflicts from using varying dependency versions, the intricacy of managing dependencies across projects, and ensuring compatibility and coherence across modules. They also stress the challenge of updating shared dependencies across multiple locations and the risk of infrastructure issues during dependency installation.

**Q15 - Hinders UI/UX consistency maintenance. (I4)** The developers express concerns about the challenges in maintaining a consistent UI/UX across different micro-frontends, citing factors such as the need for increased attention, potential difficulties in coordination between UI/UX teams and developers, and the risk of inconsistencies arising from decentralized development. Developers emphasize the importance of proactive measures such as synchronization of design efforts and the use of design systems to minimize inconsistencies but acknowledge that some level of inconsistency is likely to persist.

**Q16 - Impedes the application debugging process. (I5)** While micro-frontends don't outright impede the application debugging process, they introduce added complexities and make debugging more laborious compared to monolithic architectures. Developers note the challenge of debugging the entire application, often necessitating debugging each micro-frontend separately. They also emphasize the extra effort required to debug interactions between different micro-frontends, such as assessing the impact of a library on another micro-frontend. While some developers believe debugging individual parts of the application is manageable, they admit it's more challenging than debugging a monolithic application.

**Q17 - Increases the application's complexity. (I6)** The developers' grading of this statement varied widely. Various aspects were cited to justify their choices, including managing dependencies, interactions, deployment, testing, and maintenance. They express concerns about the growing number of micro-frontends and the challenges in managing and integrating them. They also highlight the difficulty in ensuring consistency and maintaining dependencies across multiple micro-frontends. However, some developers acknowledge that there is a trade-off. In terms of code, complexity might actually decrease, and a well-defined architecture may help mitigate some challenges.

**Q18 - Requires more coordination between the teams. (I7)** Participants recognize the importance of communication and alignment among teams to prevent conflicts and ensure the smooth functioning of different application parts. While micro-frontends may enable teams to work more independently on specific components, some developers suggest that this independence could potentially increase the need for coordination, especially for broader changes and when components require interaction. However, other participants believe that there will always be a necessity of alignment between teams, regardless of the approach.

**Q19 - Increases the fragmentation of knowledge about the application. (I8)** All developers partially agree that adopting micro-frontends can fragment knowledge about the application. This fragmentation arises from independent development and the flexibility to use various technologies. While some recognize similar fragmentation in monolithic architectures, micro-frontends tend to exacerbate this division, with teams concentrating on specific project groups. Furthermore, developers suggest that distributing responsibilities and stories among teams, combined with coordination by product owners, may alleviate this challenge.

**Q20 - Increases the risks of behaving differently on the production environment. (I9)** Participants were evenly split between grades 4 and 5. They attribute the increased risk to the complexity of the environment, dynamic component updates, and the isolated nature of the work. Although integration tests in the shell may mitigate this risk, challenges remain in accurately reflecting the production environment in testing. Additionally, factors such as dependency management and coordination during deployment contribute to the heightened risk compared to monolithic architectures.

**Q21 - Increases the chances that bugs will slip to production environment. (I10)** Developers agree that there is an increase in the likelihood of bugs slipping into the production environment. This is due to the complexity of the environment and the challenges associated with testing integration. While integration tests in the shell and team maturity in testing the application as a whole may help mitigate this risk, the potential for configuration and dependency variations complicates testing across different environments.

### 4.3 Technology Acceptance Results

According to our evaluation strategy, before the focus group, developers were assigned an exercise to enable them to explore aspects of the migrated architecture's implementation, such as dependency management and the setup of multiple micro-frontends for development. This hands-on experience with the technology was expected to enrich the subsequent focus group discussion.

Therefore, ahead of the workshop day, each participant received instructions to set up certain aspects of their local development environment for the exercise. During the session, developers were assigned to individual breakout rooms and were asked to screen share, allowing researchers to move between rooms and observe their problem-solving approaches. Participants could request assistance from researchers at any point if they encountered difficulties. Among the eight participants, two were unable to complete the exercise due to setup challenges, while the remaining six completed it with varying levels of assistance.

At the end of the workshop, to address RQ2, each participant filled out a Technology Acceptance Model (TAM) based questionnaire. Figure 6 illustrates the results. The first four questions (Q1-Q4) aimed to evaluate developers' perceptions of the architecture's usefulness, while the subsequent four (Q5-Q8) focused on its perceived ease of use. Responses to both sets of questions tended to range from neutral to positive.

The final question (Q9) assessed developers' intentions to adopt the architecture. In contrast to the generally positive or neutral responses to previous questions, this question revealed a slightly negative trend, with the majority falling within the neutral category. This suggests that while participants generally perceive the architecture as useful and moderately easy to use, they exhibit hesitance towards adopting it. The current context of the project could help to explain this hesitation. With the project divided into several applications, each with its own repository, teams already have the ability to work independently, provided they are not collaborating on the same application. Consequently, this existing setup may diminish their willingness to embrace the increased complexity of the new architecture, especially if the benefits are not perceived as particularly significant.

## 5 DISCUSSION

The synthesis of the focus group discussion revealed a nuanced understanding among developers regarding the implementation of micro-frontends. In general, the developers agree, to some extent, with most of the benefits and issues raised by Peltonen *et al.* [18].

According to what was discussed, micro-frontends offer developers flexibility in technology selection, enabling the integration of diverse technologies within the same application. This flexibility allows for tailored solutions to specific challenges, empowering development teams to leverage suitable tools for different parts of the system. At the same time, concerns were raised on the desirability to have the application composed of several different technologies, and the added complexity this may imply.

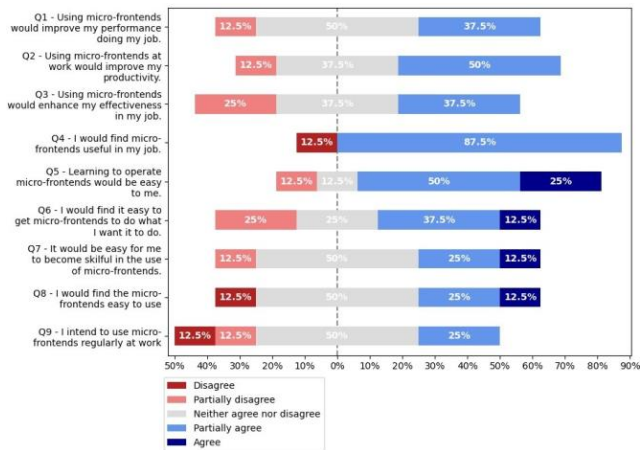

Figure 6: Technology Acceptance Model Results

Additionally, micro-frontends improve team autonomy by dividing work effectively and reducing conflicts over shared code. They also hold promise for scalability and performance, providing the framework for independent loading and development, thus potentially enhancing application scalability and user experience. Furthermore, micro-frontends offer a gradual transition path to new technologies, allowing for evaluation without the need for a complete overhaul of the existing application architecture.

However, the discussion also highlighted several challenges associated with micro-frontends. Chief among these is the complexity introduced by the architecture, spanning aspects such as dependency management, debugging, and maintaining UI/UX consistency. Effective coordination and communication between teams emerge as critical factors in navigating these challenges, underscoring the importance of alignment to prevent conflicts and ensure smooth collaboration. Testing and deployment present additional hurdles, with concerns about integration testing, deployment consistency, and discrepancies between testing and production environments. Moreover, while micro-frontends hold promise for code reusability, concerns persist regarding their effectiveness compared to monolithic architectures and the prevalence of duplicated code.

Overall, anyone intending to adopt this architecture should analyze if the gains in flexibility and scalability are worth the added complexity in infrastructure and organization. But there are other interesting points that can be taken from the discussion.

**Managing Dependencies vs Duplicating Code.** One of the primary concerns raised by the developers was the management of dependencies. This issue, typically familiar to them, stood out as a significant challenge during the discussion, signaling potential underlying complexities. An alternative approach, discussed by one developer, suggests replicating parts of the code, such as models and backend communication services, across all micro-frontends instead of relying heavily on unifying libraries. While this alternative simplifies repository management and reduces the coupling between micro-frontends, it introduces the task of maintaining duplicated code across multiple locations. Effectively handling shared code emerges as a critical challenge for improving the usability of micro-frontends architectures. Developing tools to enhance dependency version tracking or managing replicated code as a unified library could mitigate some of these issues, thereby expanding the applicability of micro-frontends architectures in various scenarios.

**Client-side Composition vs Composition During Compilation Time.** As detailed in the intervention description, an option existed to compose the application during compilation, given the absence of customization for each user. However, the decision was made to compose it on the client side, utilizing module federation, to explore new potentialities of this architecture. The former approach may have been more suitable for the context of this application, as it permits the development of each micro-frontend as a library and bundling everything as a single app. This simplifies some of the infrastructure complexities that developers viewed unfavorably, such as serving each micro-frontend at a distinct URL and managing their versions appropriately in each environment. Conversely, the chosen approach offers greater flexibility in terms of deployment independence and application customization possibilities.

**Choosing the Appropriate Split and Structuring the Application.** During the discussion, the topic of how to partition the application was frequently raised, with a perception of its critical importance and the potential costs associated with altering it later. There was a sentiment that the architecture would enhance micro-frontend reusability, particularly with a horizontal split. However, some caveats of this division were highlighted, such as the perceived increase in overall complexity and the potential limitation on team autonomy. These perceptions give rise to several questions. For example, how does the partitioning affect dependencies? In a vertical split where communication is limited to a between-page approach, could dependencies be reduced? When might it be advisable to restructure the application's use cases to adopt this split? And which aspects should be prioritized in this restructuring process?

**Integration Tests.** As our group delved deeper into web development, we became increasingly interested in end-to-end tests using Cypress[6] to automate tasks that would otherwise be manual. These tests, which simulate user interactions with the application, offer greater resilience against becoming outdated compared to traditional unit tests. Consequently, they have become our team's preferred method for preventing interface bugs. However, a significant drawback of our current approach is the extensive code duplication between different applications. This duplication likely contributed to developers' reservations about the architecture. If each micro-frontend had to replicate our current implementation, it would inflate the code base and increase maintenance costs. Moreover, these tests significantly prolong the build pipeline execution time. If the architecture leads to cascading pipelines, as seems likely to ensure reliable bug detection, it could severely disrupt the code integration process. Addressing these issues could significantly enhance the architecture's usability. Establishing guidelines for testing micro-frontends to minimize duplication and accelerate development pipelines would be advantageous. Additionally, resolving the dependency management challenge by centralizing common testing code could streamline the process further.

## 6 THREATS TO VALIDITY

This section discusses the different threats to validity identified and the corresponding mitigation actions, following the categories suggested by Wohlin *et al.* [22].

**Internal Validity.** As we implement this architectural transition, it's expected that various changes and optimizations will naturally occur, potentially confounding our ability to attribute all system modifications directly to the micro-frontends adoption. In essence, some system adjustments may be driven by general improvements or enhancements that are standard practice during ongoing development efforts. Additionally, the influence of more experienced peers in the focus group discussions might lead some developers to align with their views rather than express their own insights. Moreover, developers did not have an extensive experince with the architecture. However they were presented to the restructured application and had a practical exercise to get familiar with the new architecture before the focus group discussion. This, allied with their experience with development, allowed them to present their opinion on the matters evaluated. Indeed, the openly available board information and transcripts allow understanding that the focus group enabled valuable discussions.

**Construct Validity.** It is essential to consider the construct validity within the framework of our experiment, as the implementation of a micro-frontends architecture can vary significantly. In our specific experiment, the choices made in adopting this architecture were tailored to align with the unique context of the software under examination. Consequently, it's important to recognize that the conclusions drawn from our experiment may not universally apply to all projects utilizing micro-frontends. The context-specific nature of our decisions should be kept in mind when interpreting our findings. Notably, the influence of the choice of division of shared dependencies was particularly evident among participants. The intervention decisions were internally discussed within the team of authors, including researchers active with MFE research and practice. Hence, we believe that, while the decisions are specific, they are a representative instance for decisions typically taken when decomposing software projects into MFEs. Moreover, a potential source of bias arises from the involvement of the intervener, who is a member of the development team. This proximity to the project could inadvertently influence data collection, particularly during interviews and focus group discussions.

**Conclusion Validity.** The scope of the intervention is inherently bound by the number of developers associated with the project. Consequently, there are limitations to the inclusion of additional subjects. However, it is important to note that the evaluation of the intervention involves a multi-faceted approach. This includes a presentation and group discussion conducted with all subjects concurrently, ensuring consistency and minimizing potential variations in treatment delivery.

**External Validity.** In alignment with the principles of conventional action research, it is imperative to recognize the inherent limitations pertaining to external validity. Given that the intervention is tailored to a specific industrial case, and the structure of the focus group may introduce bias, we exercise caution in making broad claims about the generalizability of our findings to other contexts. However, it is worth emphasizing that our research unfolds within an industrial environment. The empirical observations and insights derived from this context can serve as valuable catalysts for generating hypotheses. These hypotheses, when subjected to rigorous testing and replication in diverse settings, can contribute to a more comprehensive understanding of the broader applicability and potential variations of our findings.

## 7 CONCLUDING REMARKS

This paper investigated the migration of an industry project to a micro-frontends architecture. Based on the data generated by the focus group conducted during the evaluation step, we were able to further explore the benefits and issues raised by Peltonen *et al.* [18]. This exploration contributes to the current empirical evidence with insights from industry practitioners, helping to provide a nuanced understanding.

We provide technical details of the intervention, which comprises a specific migration to a micro-frontends architecture. Software teams embarking on the journey towards micro-frontends would benefit from the reported technical considerations, organizational challenges, and implementation strategies, offering actionable insights for software development teams.

As an outcome of our evaluation, besides the architectural decision rationale and in-depth discussions of the benefits and limitations, we conclude that, while micro-frontends is a promising technology, unresolved issues (e.g., complexity of managing dependencies) still limit their broader applicability. Hence, developers generally perceived the architecture as useful and moderately easy to use but exhibited hesitance toward adopting it.

Future research should address the limitations of this study by exploring a wider array of implementation approaches and involving additional projects and developers. Also, while running experiments in professional contexts is extremely costly and difficult, simulating the maintenance task on both the monolithic and micro-frontends versions of the application could allow for a more direct, quantitative comparison of both architectures. Moreover, efforts to streamline development using the architecture and addressing issues like dependency management and integration tests could enhance the technology's acceptance.

## ONLINE RESOURCES

Our data, artifacts, and additional resources are openly available at Zenodo [2] under Creative Commons Attribution license.

## ACKNOWLEDGMENTS

We thank all the participants of the Tecgraf Institute and the Brazilian Council for Scientific and Technological Development (CNPq process #312275/2023-4) for the financial support.